\documentstyle[12pt]{article} 
\author{L.~Didukh \\
{\small \it Ternopil State Technical University, Department of Physics,}\\ 
{\small \it 56 Rus'ka Str., Ternopil UA--282001, Ukraine; 
E-mail: didukh@tu.edu.te.ua}}

\date{}
\title{A modified form of the polar model of crystals}

\begin{document}
\maketitle

\begin{abstract}
A model for description of materials with narrow energy bands is proposed.
It is shown that in narrow-band materials electron-hole symmetry is absent
in contrast to the Hubbard model. In this paper metal-insulator transition
is studied. The obtained results are compared with experimental data for 
narrow-band materials. Some specific narrow-band effects are discussed.

PACS number(s): 71.28.+d, 71.27.+a, 71.10.Fd, 71.30.+h
\end{abstract}

\section{Introduction}

1. The fact that amasing properties of narrow-band systems 
(for example oxides, sulphides and selenides of transition matals) 
are caused by electron-electron interactions is generally accepted
in our time.
But in spite of the great number of the papers concerning this problem
it is still an actual problem of condensed matter physics to construct
the consistent theory of narrow-band systems.
During last years a range of problems connected  with 
corellation problem and a necessity of an investigation of narrow band systems
have been greatly increased because of discovering of high-$T_c$ 
superconductors.  

Arising problems can be separated on three groups:
1) a construction of the narrow-band systems models using adequate 
 Hamiltonians;
2) an elaboration of effective mathematical methods to study
the model Hamiltonians;
3) a construction of the consistent theory of correlation effects
 and an explaination of peculiarities  of  physical properties
in the narrow-band systems.

Problems 1) and 2) had been considered and partially solved by S.Shubin and
S.Wonsowsky in their famous theory of the polar model [1].
In that theory the polar model Hamiltonian and its ``configurational'' 
representation had been proposed.
The polar model prove to be very meaningful owing to 
euristic value of ``configurational'' description idea (basical
for model treatment of $3d$-compounds~[2].)
In the frames of the polar model a criterion of metal-insulator
transition (MIT) was formulated for the first time; an explanation of
fractional atom momentum in transition $3d$-metals,
a hypothesis of the possibility of charge ordering were proposed;
a possibility of being gapeless semiconductor and 
superexchange interaction were predicted.
Commonly used the Hubbard model is a partial case of the polar model~[3].

However, a direct use the polar model (in traditional form~[1])
to solve the problems had been proved to be not effective in many cases.

Firstly, the transition from the second quantization Hamiltonian in 
terms of electron operators to its representation in terms of 
Shubin-Wonsowsky operators had been realized by the substitution of
some combination of electron operators through the combination
of  Shubin-Wonsowsky operators  with the same action on wave function.
Such transition is bulky and difficult even for $s$-band (see for example~[4]). 

Secondly, the approximations underlied mathematical treatment of the polar
model are uncontrollable (first of all, the postulation of Bose-type 
commutation rules for the operators of current excitation).

2. The polar model theory was developed in two ways. First is 
connected with developing the methods of effective mathematical treatment 
of initial polar model Hamiltonian in electron representation.
The fundamental results in this way belong to N.Bogolubov~[5].
He proposed the effective Hamiltonian method which took into account 
high energy  electron states with the help of special form of perturbation
theory. This method is one of the most consistent approach to study 
exchange interactions in the magnetic insulators~[6]. An use the 
configurational representation of the polar model (polar and homeopolar 
states) proved to be helpful for an interpretation of the obtained results,
a control of the performed calculations (sometimes very bulky).

 Second way is based on immediate an use configurational representations.
This approach is effective for an investigation of peculiarities of
narrow-band systems, insulators and semiconductors, on the one hand,
metals and materials in which MIT is caused by external influences,
on the other hand.   

 Important achievement in this way was obtained in the works of Lviv 
branch in theory of solids. Here works [7, 8, 9] of A.~Yu.~Glauberman,
V.~V.~Vladimirov,  I.~V.~Stasyuk were initial.
 On this way the important problems of the polar model of non-metallic
cristals (the problem of constructing model Hamiltonians in terms of 
site elementary excitations and the problem of commutation rules for
operators of site elementary excitations) were solved. 

3. Wide use the configurational forms of model Hamiltonians to 
consider physical properties of narrow-band materials is associated with
paper of J.Hubbard~[10] in which $X_i^{kl}$-operators had been 
introduced and papers~[11, 12]  where 
the relation between elecron operators and
transition-operators had been first established,
an effective form of perturbation theory had been proposed.
It had been shown that the proper identification of the Shubin-Wonsowsky 
operators and transition-operators (the Hubbard operators)
leads to formal equivalence of the traditional form of the polar model
and its modern representations in terms of transition-operators. 
It also had been showh that 
\begin{eqnarray*}
X_i^{kl}=\alpha_{ik}^{+}\alpha_{il},
\end{eqnarray*}
where $\alpha_{ik}^{+},\ \alpha_{il}$ -- Shubin-Wonsowsky operators 
of creation and destruction of $|k>$-state and $|l>$-state respectively
on $i$-site.

 An expediency of using $X_i^{kl}$- or $\alpha^{+}_{i\nu}\alpha_{i\mu}$-
representation is predicted by requirements of considered problem.
In calculations using diagrammatic Green functions technique or
perturbation theory it is convenient to use $X_i^{kl}$-representation
of Hamiltonian~[13], in problems using approximate second 
quantization method (e.g. to study MIT using the mean-field approximation 
(MFA) in generalized
Hartree-Fock approximation~[14]) $\alpha^{+}_{i\nu}\alpha_{i\mu}$-
representation is more convenient. The $b$-$c$-representation (see Sect.~2)
also can be useful.

4. The representations of model narrow-band Hamiltonians in terms of
$X_i^{kl}$-, $b$-$c$- or $\alpha^{+}_{i\nu}\alpha_{i\mu}$-operators are 
helpful to
understanding physics of correlation effects in narrow energy bands, 
to explain physical properties of narrow-band materials.
These representations are convenient from point of view of 
mathematical treatment of models. 

Below a consistent form of polar model of narrow-band materials is proposed
and consequences of this model are considered.

\section{The Hamiltonian}
\setcounter{equation}{0}

Hamiltonian of the system of $s$-electrons in the Wannier-representation
is written as
\begin{eqnarray}
H&=&-\mu \sum_{i\sigma}a_{i\sigma}^{+}a_{i\sigma}+
{\sum}'_{ij\sigma}t(ij)a_{i\sigma}^{+}a_{j\sigma}+
\\
\nonumber
&+&{1\over 2} \sum_{\stackrel{ijkl}{\sigma,\sigma^{'}}}J(ijkl)
a_{i\sigma}^{+}a_{j\sigma^{'}}a_{l\sigma^{'}}a_{k\sigma},
\end{eqnarray}
where $a_{i\sigma}^{+}$, $a_{i\sigma}$ --  creation
and destruction operators of electron on site $i$, 
$\sigma =\uparrow, \downarrow$, $\mu$ -- chemical potential,
 $n_{i\sigma}=a_{i\sigma}^{+}a_{i\sigma}$,
\begin{eqnarray}
&& t(ij)=\int{\phi}^{*}({\bf r}-{\bf R}_{i})\sum_{n}V({\bf r}-{\bf 
 R}_{n})\phi({\bf r}-{\bf R}_{j})d{\bf r},
\\
&& J(ijkl)=\int{\int{\phi}^{*}}({\bf r}-{\bf R}_{i})\phi({\bf r}-{\bf R}_{k}){e^{2}\over |{\bf r}-{\bf r}^{'}|}\times
\\
\nonumber
&& \times\phi^{*}({\bf r}^{'}-{\bf R}_{j})\phi({\bf r}^{'}-{\bf R}_{l})d{\bf r}d{\bf r}^{'},
\end{eqnarray}
 are the matrix elements which describe hoppings of electrons
between nearest-neighbour sites of lattice in consequence of electron-ion 
($V({\bf r}-{\bf R}_{i})$ -- potential energy of electron interacting
with an ion on $i$-site)
and electron-electron interactions. The prime by second sum in 
Eq.~(2.1) signifies that $i\neq{j}$. 

Narrow energy bands allow to simplify Hamiltonian~(2.1). 
Here wave functions are closely resemble atomic $3d$-functions (their 
overlapping decrease quickly with increase of the inter-atomic spacing),
so matrix elements $t(ij)$ and $J(ijkl)$ can be estimated from
degree of overlapping. Thus quantities 
$J(iiii)$ and $J(ikik)$ will be of zero ®rder,
$J(iiij)$, $J(ijkj)$ -- of first order (as $t(ij)$), $J(ijkl)$
at $i\neq{k},\ j\neq{l}$ -- of second order (immediate estimation of
$J(ijkl)$ is given in the paper~[3]). In accordance with this we 
limit ourselves to accounting in Hamiltonian~(2.1) matrix
elements $J(iiii)=U$, $J(ijij)=V(ij)$ ($÷$ and $j$ are nearest neighbours),
$J(iiij)=T(ij)$, $J(ijkj)$ ($k\neq{i},\ k\neq{j}$),
$J(ijji)=J(ij)$; taking into account quantity of second order $J(ij)$ 
is on principle necessary to describe ferromagnetism in this model
in Mott-Hubbard insulator state. 'hen
\begin{eqnarray}
H=&-&\mu \sum_{i\sigma}a_{i\sigma}^{+}a_{i\sigma}+\\ \nonumber
&+&{\sum}'_{ij\sigma}a_{i\sigma}^{+}(t(ij)+
\sum_{k}J(ikjk)n_{k})a_{j\sigma}+
U\sum_{i}n_{i\uparrow}n_{i\downarrow}+
\\
&+&{1\over 2}{\sum}'_{ij\sigma \sigma^{'}}J(ij)a_{i\sigma}^{+}
a_{j\sigma{'}}^{+}a_{i\sigma{'}}a_{j\sigma}+
{1\over2}{\sum}'_{ij\sigma\sigma{'}}V(ij)n_{i\sigma}n_{j\sigma{'}},
\nonumber
\end{eqnarray}
where $n_{i}=n_{i\uparrow}+n_{i\downarrow}$.

In Hamiltonian~(2.4) we rewrite the sum 
$\sum'_{ij\sigma k}J(ikjk)a^{+}_{i\sigma}n_ka_{j\sigma}$
in the form
\begin{eqnarray}
{\sum_{ij\sigma}}'\sum_{\stackrel{k\neq{i}}{k\neq{j}}}J(ikjk)a^{+}_{i\sigma}
n_ka_{j\sigma}+{\sum_{ij\sigma}}'\left(J(iiij)a^{+}_{i\sigma}a_{j\sigma}
n_{i{\bar \sigma}}+h.c.\right)
\end{eqnarray}
(${\bar \sigma}$ denotes spin projection which is opposite to $\sigma$).
We suppose (as in papers~[14, 15]) that
\begin{eqnarray*}
{\sum_{ij\sigma}}'\sum_{\stackrel{k\neq{i}}{k\neq{j}}}J(ikjk)a^{+}_{i\sigma}
n_ka_{j\sigma}=n\sum_{\stackrel{k\neq{i}}{k\neq{j}}}J(ikjk)
{\sum_{ij\sigma}}'a^{+}_{i\sigma}a_{j\sigma}
\end{eqnarray*}
with $n=\langle n_{i\uparrow}+n_{i\downarrow}\rangle$ (sites $i$ and $j$ 
are nearest neighbours). It should be noted that this supposition is exact in
the homeopolar limit ($n_i=1$).

Thus Hamiltonian~(2.4) takes the following form
\begin{eqnarray}
H=&-&\mu \sum_{i\sigma}a_{i\sigma}^{+}a_{i\sigma}+
{\sum}'_{ij\sigma}t_{ij}(n)a_{i\sigma}^{+}a_{j\sigma}+\\ \nonumber
&+&{\sum}'_{ij\sigma}\left(T(ij)a_{i\sigma}^{+}a_{j\sigma}n_{i\bar{\sigma}}
+h.c.\right)+U\sum_{i}n_{i\uparrow}n_{i\downarrow}+
\\
&+&{1\over 2}{\sum}'_{ij\sigma \sigma^{'}}J(ij)a_{i\sigma}^{+}
a_{j\sigma{'}}^{+}a_{i\sigma{'}}a_{j\sigma}+
{1\over2}{\sum}'_{ij\sigma\sigma{'}}V(ij)n_{i\sigma}n_{j\sigma{'}},
\nonumber
\end{eqnarray}
where
\begin{eqnarray}
t_{ij}(n)=t(ij)+n\sum_{\stackrel{k\neq{i}}{k\neq{j}}}J(ikjk)
\end{eqnarray}
is the effective hopping integral between nearest neighbours.

Neglecting all matrix elements in~(2.6) except $t(ij)$ and $J(iiii)$
we obtain the Hubbard Hamiltonian.

The transition from the general form of Hamiltonian~(2.6) to the Hubbard 
Hamiltonian, i.e. taking into account only intra-atomic Coulomb repulsion,
usualy is argued by smallness of quantities $J(iiij)$,
$J(ikjk)$, $J(ijji)$ and $J(ijij)$ in comparison  with $J(iiii)$. 
However, taking into account these matrix elements can be on principle
important from a points of view of both a construction of correlation effects
theory in materials with narrow energy bands and  an interpretation of
physical properties of these materials~[13, 15-17].

Neglecting inter-atomic exchange interaction is justified by a smallness
of $J(ij)$ in comparison with $U$ and hopping integral $t(ij)$, on the one
hand, a possibility of ferromagnetic ordering stabilization in narrow
energy band (NEB) in consequence of ``translational'' mechanism of exchange, 
on the other hand. Without consideration of a possibility of ferromagnetism 
realization in the one-band Hubbard model it should be noted that in NEB
a contribution of translational part of energy in total system energy
can be smaller then a contribution of energy of inter-atomic exchange
interaction (in spite of the fact that $|t(ij)|>>J(ij)$). 
Really, in partially filled NEB (for $U>>|t(ij)|$)
the contribution of translational part of ground state energy
$\sim n\delta w$ ($\delta$ -- degree of deviation from half-filling,
$n$ is the average number of electrons on site, $2w$ is the bandwidth)~[16], 
and the contribution of exchange interaction in ground state energy
$\sim zn^{2}J$
($J$ -- exchange integral between nearest neighbours, $z$ is the
number of nearest neighbours to a site). It is clearly, that in NEB close
to half-filling ($\delta \to 0$) the contribution of energy of 
inter-atomic exchange interaction in total system energy will be 
biggest.
In particular, in non-doped Mott-Hubbard ferromagnets magnetic ordering
is stabilized by inter-atomic exchange interaction only.

Taking into account inter-atomic Coulomb interaction is on principle
important to understand a character of charge ordering in meterials with
NEB.

Finally, neglecting correlated hopping~(2.5) is justified by the estimation of
matrix elements~[3]. However, matrix elements $J(ikjk)$ are hopping
integrals. Thus taking into account~(2.5) leads to the renormalization of
translational processes describing band part of Hamiltonian~(2.6). At this
$t(ij)$, $T(ij)$, $J(ikjk)$  are quantities of the same order.

If direct exchange interaction and inter-atomic Coulomb repulsion
can be taken into account by respective renormalization of chemical
potential (ferromagnetic and charge orderings are absent) then
Hamiltonian~(2.6) takes the form  
\begin{eqnarray}
H=&-&\mu \sum_{i\sigma}n_{i\sigma}+
{\sum}'_{ij\sigma}t_{ij}(n)a_{i\sigma}^{+}a_{j\sigma}+\\
&+&{\sum}'_{ij\sigma}\left(T(ij)a_{i\sigma}^{+}a_{j\sigma}
n_{i\bar{\sigma}}+h.c.\right)+
U\sum_{i}n_{i\uparrow}n_{i\downarrow}.
\nonumber
\end{eqnarray}

As have been noted the peculiarity of the model of material with NEB
describing Hamiltonian~(2.8) is taking into account (as on principle 
important) inter-site hoppings of electrons which are caused by electron-
electron interaction and inter-atomic Coulomb and exchange interactions.
In this connection the following fact should be noted. 
Formally, correlated hopping had been introduced in some papers begining
from the pioneer work of S.~Shubin and S.~Wonsowsky~[1]; on the possible
renormalization of ``band'' hopping in consequence of taking into account
correlated hopping had been noted in the papers~[5, 18, 19]. First, the
important role of correlated hopping in NEB when a description is adequate
with the help of analogues of the Hubbard subbands had been pointed out
in the paper~[20]. In that work, in particular, it had been shown that NEB 
had electron-hole asymmetry and essentially renormalized bandwidths
connected by hopping in ``hole and doublon subband''. This approach has 
been developed in the papers~[21, 22], where it has been shown that
some properties of narrow-band materials can be interpreted using the
idea of correlated hopping and caused by it electron-hole asymmetry in NEB.

On the fact that taking into account correlated hopping is of principle
necessary was pointed out also in the papers~[23]. In recent years models with
correlated hopping have been studied intensively~[24--28].

\section{Partial cases of polar model} 
\setcounter{equation}{0}

\subsection{Weak intra-atomic interaction}

To simplify the consideration we use model Hamiltonian~(2.8). If 
intra-atomic Coulomb interaction is weak ($U<|t_{ij}(n)|$) 
then we can take into account 
the electron-electron interaction in the Hartree-Fock approximation:
\begin{eqnarray}
&&n_{i\uparrow}n_{i\downarrow}=n_{\uparrow}n_{i\downarrow}+
n_{\downarrow}n_{i\uparrow},
\\
\nonumber
&&a_{i\sigma}^{+}n_{i\bar{\sigma}}a_{j\sigma}=n_{\bar{\sigma}}
a_{i\sigma}^{+}a_{j\sigma}+
\langle a_{i\sigma}^{+}a_{j\sigma}\rangle n_{i\bar{\sigma}},
\end{eqnarray}
where the average values $\langle n_{i\sigma}\rangle=n_{\sigma}$ 
are independant
of site number (we suppose that distributions  of electron charge and 
magnetic momentum are homogenous). Taking into account~(3.1) we can write 
Hamiltonian~(2.8) in the following form:
\begin{eqnarray}
&& 
H={\sum}'_{ij\sigma}\epsilon_\sigma(ij)a_{i\sigma}^{+}a_{j\sigma},
\end{eqnarray}
where
\begin{eqnarray}
&& \epsilon_\sigma(ij)=-\mu+\beta_\sigma+n_{\bar{\sigma}}U +
t_{ij}(n\sigma);
\end{eqnarray}

\begin{eqnarray}
&& \beta_\sigma={2 \over N}\sum_{ij} T(ij) \langle 
a_{i\bar{\sigma}}^{+}a_{j\bar{\sigma}}\rangle,
\end{eqnarray}

\begin{eqnarray}
&& t_{ij}(n\sigma)=t_{ij}(n)+2n_{\bar{\sigma}}T(ij).
\end{eqnarray}
The dependences of effective hopping integral on electron concentration 
and magnetization, a being of the spin-dependent displasement of band
center are the essential distinction of single-particle energy spectrum
in the model described by Hamiltonian~(3.2) from the spectrum in the 
Hubbard model for weak interaction. An use of~(3.2) allows, in particular,
to explain the peculiarities of dependence of binding energy on 
atomic number in transition metals and also essentially modifies
theory of ferromagnetism in a collective electron model.

\subsection{ Strong intra-atomic interaction}

For typical narrow-band materials the conditions of strong $U>>t(ij)$ 
or moderate $U \sim{t(ij)}$ intra-atomic Coulomb repulsion are satisfied. 
In this case Hamiltonian~(2.6) using ``configurational ideology'' of 
the polar model 
proposed in~[12] can be written in form suitable for mathematical treatment.
Let us rewtrite Hamiltonian~(2.6) in ``configurational'' 
representations~[11, 12].
Transitions to $\alpha$-operators are given by the formulae:
\begin{eqnarray}
\nonumber
a_{i\sigma}^{+} = \alpha_{i \sigma}^{+}\alpha_{i 0} -
\eta_{\sigma} \alpha_{i 2}^{+} \alpha_{i \bar{\sigma} }, \ \
a_{i\sigma} = \alpha_{i 0}^{+} \alpha_{i \sigma} -
\eta_{\sigma} \alpha_{i \bar{\sigma}}^{+} \alpha_{i 2},
\end{eqnarray}
where
$\eta_{\sigma}=+1$ when $\sigma = \uparrow$,
$\eta_{\sigma}=-1$ when $\sigma = \downarrow$, 
site $i$ can be not occupied with electron ($|0 \rangle$), singly occupied 
($| \sigma \rangle$) or doubly occupied ($|2 \rangle$).
Transitions to $X$-operators are given by the formulae:
\begin{eqnarray}
a_{i\sigma}^{+}=X_i^{\sigma 0}-\eta_{\sigma} X_i^{2\bar{\sigma} }, \ \
a_{i\sigma}=X_i^{0\sigma}-\eta_{\sigma} X_i^{\bar{\sigma}2},
\end{eqnarray}
where $X_i^{kl}$ -- operators of site $i$ transition from state 
$|l\rangle$ to state $|k\rangle$,
\footnote{In the papers~[11, 12] notations of site transition-operators	
$B_{kl}^i$ had been introduced. In this paper we use modern notations 
$X^{kl}_i$, and more convenient notations of state $|ik\rangle$.}

The Hamiltonian can be written as:
\begin{eqnarray}
H=H_0+H_1+H'_1+H_{ex},
\end{eqnarray}
where
\begin{eqnarray}
&& 
H_0=-\mu\sum_i\left(X_i^{\uparrow}+X_i^{\downarrow}+2X_i^2\right)+
U\sum_{i}X_i^2+\\
\nonumber
&&\null\quad+{1\over 2}\sum_{ij}V(ij)\left(1-X_i^0+X_i^2\right)\left(1-X_j^0+X_j^2\right),\\
&& H_1={\sum}'_{ij\sigma}t_{ij}(n)X_i^{\sigma 0} X_j^{0\sigma} + 
\sum_{ij\sigma}\tilde{t}_{ij}(n)X_i^{2\sigma}X_j^{\sigma 2},
\\
&& H'_1={\sum}'_{ij\sigma}\left(t'_{ij}(n)
\left(X_i^{\downarrow 0}X_j^{\uparrow 2}-X_i^{\uparrow 0}X_j^{\downarrow 2}\right)+h.c.\right),
\\
&& H_{ex}=-{1\over 2}{\sum\limits_{ij\sigma}}'J(ij)\left(
\left(X_i^{\sigma}\!+\!X_i^{2}\right)\left(X_j^{\sigma}\!+\!X_j^{2}\right)+
X_i^{\sigma\bar{\sigma}}X_j^{\sigma\bar{\sigma}}\right);
\end{eqnarray}
$X_i^k$  is the operator of number of $|k\rangle$-states on site $i$,
\begin{eqnarray}
\tilde{t}_{ij}(n)=t_{ij}(n)+2T(ij),
\\
t'_{ij}(n)=t_{ij}(n)+T(ij).
\end{eqnarray}

Expedience of configurational representation is proved by
the fact that intra-atomic interaction takes the diagonal form.
Besides effects of intra-atomic Coulomb interactions correlating
electron translations are described by Hamiltonians $H_1$ and $H'_1$.

$H_1$ describes transitions of $|j\sigma\rangle$-configurations to 
$|i0\rangle$-configurations and{${|j\uparrow\downarrow\rangle}$}-
configurations to $|j\sigma\rangle$-configurations of neighbour sites
which forms $\sigma-0$--subband --``hole'' subband and
$2-\sigma$--subband --``doublon''subband respectively (which are analogues
of ``lower'' and ``upper''Hubbard subbands).

$H'_1$ describes transitions between $\sigma-0$- and 
$\uparrow\downarrow-\sigma$-subbands ( processes of paired creation
and destruction of holes and doublons). These processes are
``translational'' in the distinction from ``activational''processes
described by $H_1$.

 If we neglect inter-atomic Coulomb  and exchange interaction
in Hamiltonian~(3.7) then the Hamiltonian takes the operator structure
equivalent to the Hubbard Hamiltonian one. However in this model
hopping integrals in $\sigma-0$-- and 
$\uparrow\downarrow-\sigma$--subbands and ``interbands''
hopping integrals are dependent on concentration and different,
as distinction from the Hubbard model (see Fig.~1). 
Properties of this ``asymmetrical Hubbard model'' can be essentially 
different. 

\subsection{Generalized $t-J$ model}

Configurational representation is especially useful in an investigation of 
narrow-band system in which the condition $U>>t(ij)$ is satisfied.
In this case system can be both Mott-Hubbard insulator at $n = 1$
and  doped Mott-Hubbard insulator at $n \neq{1} $. Then general
Hamiltonian using suitable form of perturbation theory~[12] 
generalizing Bogolubov perturbation theory~[5] for metallic systems
can be written in the form of effective Hamiltonian which is convenient 
for the mathematical treatment. Thus trasition to the well-known 
$t-J$ model occurs (see the review~[29] and also the papers~[12, 16]
where modern form of $t-J$-model was formulated first).
Let us use the approach proposed in~[12] for generalized narrow-band 
Hamiltonian~(3.7). Perform the canonical transformation 
\begin{eqnarray}
\tilde{H}=e^sHe^{-s},
\end{eqnarray}
where
\begin{eqnarray}
S=\sum_{ij}\left(L(ij)\left(X_i^{\uparrow 0}X_j^{\downarrow 2}
-X_j^{\downarrow 0}X_i^{\uparrow 2}\right)-h.c.\right).
\end{eqnarray}

If we limit ourselves to quantities of second order of smallness
in Eq.~(3.14)  ($S$ is of first order), then
\begin{eqnarray}
\nonumber
\tilde{H}&=&H_0+H_1+H'_1+[SH_0]+
\\
&+&[SH_1]+[SH'_1]+{1\over 2}\left[S[SH_0]\right].
\end{eqnarray}
Use the condition of an elimination of ``activational''processes
\begin{eqnarray}
H'_1+[SH_0]=0.
\end{eqnarray}
Taking into account inter-atomic Coulomb interaction in the mean-field 
approximation we obtain that
\begin{eqnarray}
L(ij)=t'_{ij}(n)/\Delta,
\end{eqnarray}
where
\begin{eqnarray}
\Delta=U-V+zV\left(\langle X_i^0 \rangle+\langle X_i^2 \rangle \right)
\end{eqnarray}
 is the activation energy of hole-doublon pair ($V$ is the strentgh of
Coulomb repulsion between nearest neighbours ).

 The components of commutator $[S;H_1]$ have operator structures similar to
structure of $H'$, but with ``hopping integrals'' of second order;
in the considered approximation they do not contribute to $\tilde{H}$.
Thus for the case of $\sigma$-0- and $\uparrow\downarrow$-$\sigma$-subbands
are separated by energy gap and $t'_ij(n)<<\Delta$ the initial 
Hamiltonian~(2.6) has the form
\begin{eqnarray}
\nonumber
\tilde{H}&=&H_0+{\sum}'_{ij}t_{ij}(n)X_i^{\sigma 0}X_j^{0\sigma}+\\
&+&{\sum}'_{ij\sigma}\tilde{t}_{ij}(n)X_i^{2\sigma}X_j^{\sigma2}+H_{ex}+
\tilde{H}_{ex}+\tilde{H}_t,
\end{eqnarray}
where
\begin{eqnarray}
\nonumber
\tilde{H}_{ex}&=&-{1\over 2}{\sum}'_{ij\sigma}\tilde{J}(ij) (X_i^{\sigma}X_j^{\bar{\sigma}}-
\\
&-&X_i^{\sigma\bar{\sigma}}X_j^{\bar{\sigma}\sigma}-X_i^{2}X_j^{0}),
\end{eqnarray}
\begin{eqnarray}
\tilde{H}_t=-{1\over 2}{\sum}'_{ijk\sigma}J(ijk)\left(X_i^{\sigma 0}X_j^{\bar{\sigma}}X_k^{0\sigma}-
X_i^{\sigma 0}X_j^{\bar{\sigma}\sigma}X_k^{0\bar{\sigma}}\right)\\ 
\nonumber
-{1\over 2}{\sum}'_{ijk\sigma}J(ijk)\left(X_i^{2\sigma}X_j^{\sigma\bar{\sigma}}X_k^{\bar{\sigma}2}-
X_i^{2\sigma}X_j^{\bar{\sigma}}X_k^{\sigma2}\right).
\end{eqnarray}
Here
\begin{eqnarray}
\tilde{J}(ij)=2t'_{ij}(n)t'_{ij}(n)/\Delta
\end{eqnarray}
 -- integral of indirect exchange (through polar states),
\begin{eqnarray}
J(ijk)=2t'_{ij}(n)t'_{jk}(n)/\Delta
\end{eqnarray}
 -- integral of indirect charge transfer in 
$\sigma$-0- and $\uparrow\downarrow$-$\sigma$-subbands;
in sum~(3.22) sites $i$ and $k$ are nearest neighbours to $j$.

An elimination of processes of paired creation and destruction of holes
and doublons (in first order on hopping integral $t'_{ij}(n)$) leads to
a rise of two terms $\tilde{H}_{ex}$ and $\tilde{H}_t$ in EH~(3.20).
$\tilde{H}_{ex}$ describes indirect exchange interaction (superexchange),
$\tilde{H}_t$ describes indirect hopping of electrons (supperhopping).
EH~(3.20) generalizes the EH obtained in [12] for the Hubbard model.
The distinctions of EH~(3.20) from the forms of $t-J$-models
([30, 31]) are caused by the concentration-dependence of hopping integrals
in $\sigma$-0- and $2-\sigma$-subbands, firstly, the difference of the 
noted hopping integrals (the absence of electron-hole symmetry), secondly,
unusual form of the superexchange and superhopping integrals (the being
of the concentration-dependence in hopping integrals, formula~(3.19) for
$\Delta$),thirdly. 

In the modified in this way $t-J$-model, in particular, the conditions of 
a realization of high-$T_c$ are more favourable than
in the similar Spalek model~[32]. The enumerated peculiarities of the 
model EH are useful to interpret physical properties of narrow-band
materials.

\subsection{``$b$-$c$''-representation}
 We give here another one form of Hamiltonian representation
which is useful to study metal-insulator transition problem.
For simplicity we consider the Hubbard Hamiltonian. 
Let us introduce operators 
\begin{eqnarray}
b_{i \sigma}^{+}=a_{i \sigma}^{+}(1-n_{i \bar{\sigma}}),\quad
c_{i \sigma}^{+}=a_{i \sigma}^{+}n_{i \bar{\sigma}}.
\end{eqnarray}
One can see that 
\begin{eqnarray}
a_{i \sigma}^{+}=b_{i \sigma}^{+}+c_{i \sigma}^{+},\quad
a_{i \sigma}=b_{i \sigma}+c_{i \sigma}.
\end{eqnarray}
The Hubbard Hamiltonian in this ``$b$-$c$'' representation has the form
\begin{eqnarray}
H=H_0+H_1+H'_1,
\end{eqnarray}
with
\begin{eqnarray}
&& 
H_0=-\mu\sum_{i \sigma}\left(b_{i \sigma}^{+}b_{i \sigma}+
c_{i \sigma}^{+}c_{i \sigma} \right)+
{U \over 2}\sum_{i \sigma} c_{i \sigma}^{+}c_{i \sigma},
\\
&& H_1=t{\sum}'_{ij\sigma}\left(b_{i \sigma}^{+}b_{j \sigma} + 
c_{i \sigma}^{+}c_{j \sigma} \right),
\\
&& H'_1=t{\sum}'_{ij\sigma} (b_{i \sigma}^{+}c_{j \sigma} + 
h.c.),
\end{eqnarray}
where
$t \equiv t_{ij}$.
From the operator structures of latter Hamiltonians one can see that 
$H_1$ describes translational hopping forming ``$b$''- and ``$c$''-bands,
$H'_1$ describes ``inter-band'' hopping. Here one can see formal analogy 
between model described by Hamitontan~(3.7) and ``two-configuration'' 
Irkhin model~[33].

\section{Single-particle energy spectrum. Metal--insulator transition.}
\setcounter{equation}{0}

 Beyond the frameworks of approximations considered in Sect.~3
remain the region of parameters, at which a width of unperturbed band
$2z|t(ij)|$ and a strength of Coulomb repulsion are close to each
other. From general physical considerations in this region we have to
expect the metal-insulator transition (we mean $n=1$).
Although great number of the parers are devoted to the solving of energy
gap problem the question of correct description of metal-insulator
transition remain the attention of researchers (see, for example [34, 35]).

 The most significant  defect of the approximation ``Hubbard-I'' is the 
inability of a description of the metal-insulator transition (MIT) 
because of the presence of an energy gap in a spectrum  at all
values of $U/w > 0$. Other approximations  are free from  this defect,
but have their own defects~[34, 35]. 

 In this paper a new approach to calculating 
the single-particle energy spectrum of narrow-band materials which
leads to correct description of metal-insulator transition is proposed.
This approach is based on some variant of an approximate second 
quantized representation~[36] method in a 
generalized Hartree-Fock approximation (GHFA)~[37]. 

We start from the Hamiltonian~(3.27) in the ``$b-c$''-representation~(3.25).
Suppose that any kind of electron ordering is absent (in this case
taking into account the inter-ratomic interaction in the mean-field 
approximation leads to chemical potential renormalization).
Let us introduce the Green function:
\begin{eqnarray}
G_{ps}^{\sigma}(E)= \langle \langle a_{p\sigma}|a_{s\sigma}^{+}
\rangle \rangle.
\end{eqnarray}

The single-particle Green function which is written in $b$-$c$-operators as
\begin{eqnarray}
G_{ps}^{\uparrow}(E)=
\langle\langle b_{p\uparrow}|b_{s\uparrow}^{+} \rangle\rangle +
\langle\langle b_{p\uparrow}|c_{s\uparrow}^{+} \rangle\rangle +
\langle\langle c_{p\uparrow}|b_{s\uparrow}^{+} \rangle\rangle +
\langle\langle c_{p\uparrow}|c_{s\uparrow}^{+} \rangle\rangle 
\end{eqnarray}
is given by equation
\begin{eqnarray}
(E+\mu)\langle\langle b_{p\uparrow}|b_{s\uparrow}^{+} \rangle\rangle=
&&{\delta_{ps}\over 2\pi}\langle 1- n_{p {\downarrow}}\rangle+ 
\langle\langle{\left[b_{p\uparrow}, H_1\right]}_{-}|
b_{s\uparrow}^{+}\rangle\rangle
\nonumber\\
&&+\langle\langle{\left[b_{p\uparrow}, H'_1\right]}_{-}|
b_{s\uparrow}^{+}\rangle
\rangle,
\end{eqnarray} 
with ${[A, B]}_{-}=AB-BA$. Suppose in Eq.~(4.3) that
\begin{eqnarray}
{\left[b_{p\uparrow}, H_1\right]}_{-}=\sum_{j}\epsilon(pj)b_{j\uparrow},
\quad 
{\left[b_{p\uparrow}, H'_1\right]}_{-}=\sum_{j}\epsilon_1(pj)
c_{j\uparrow},
\end{eqnarray}
where $\epsilon(pj)$ and $\epsilon_1(pj)$ are 
non-operator expressions. It gives the closed system of equations for Green
functions $\langle\langle b_{p\uparrow}|b_{s\uparrow}^{+}\rangle\rangle$
and $\langle\langle c_{p\uparrow}|c_{s\uparrow}^{+}\rangle\rangle$. After
anticommutation of both sides of the first of formulae~(4.4) with 
$b_{k\uparrow}^{+}$ we have
\begin{eqnarray}
\epsilon(pk)\langle 1- n_{k {\downarrow}}\rangle=
&&t\langle 1- n_{{p \downarrow}}\rangle
\langle 1- n_{{k \downarrow}}\rangle
+tb_{p\downarrow}^{+}b_{p\uparrow}
b_{k\uparrow}^{+}b_{k\downarrow}
\nonumber\\
&&-\delta_{pk}t\sum_{j}
b_{k\uparrow}^{+}b_{j\uparrow}+
\delta_{pk}t\sum_{j}
c_{j\uparrow}^{+}c_{k\uparrow}+
tb_{p\downarrow}c_{p\uparrow}
c_{k\uparrow}^{+}b_{k\downarrow}^{+}.
\end{eqnarray}

An usual method of determination of $\epsilon(pk)$ consists
in an averaging of expression~(4.5). In this way the 
approximations~[3, 30, 38] 
are obtained; the defects of these approximations are known (see,
for example~[39]). Here, another approach is proposed (see also~[36]).

In~(4.5) we write 
\begin{eqnarray*}
b_{k\uparrow}^{+}=\alpha_{k\uparrow}^{+}\alpha_{k0},\quad
b_{k\uparrow}=\alpha_{k0}^{+}\alpha_{k\uparrow},\\
c_{k\downarrow}^{+}=-\alpha_{k\downarrow}^{+}\alpha_{k0},\quad
c_{k\downarrow}=-\alpha_{k0}^{+}\alpha_{k\downarrow},
\end{eqnarray*}
where $\alpha_{ik}^{+}$ is the creation operator of $|k\rangle$-state on 
$i$-site and $\alpha_{il}$ is the annihilation operator of $|l\rangle$-state
(analogues of Shubin-Wonsowsky opperators). 
Let us substitute $\alpha$-operators through $c$-numbers in the Eq.~(4.5)
\begin{eqnarray}
\alpha_{i\sigma}^{+}=\alpha_{i\sigma}=\left({1-2d\over 2}\right)^{1/2},
\qquad
\alpha_{i 0}^{+}=\alpha_{i 0}=\alpha_{i 2}^{+}=\alpha_{i 2}=d^{1/2}
\end{eqnarray}
(non-magnetic case, electron concentration on site $n=1$); $d$ is the 
concentration
of polar states (holes or doublons). From Eq.~(4.5) after transition to 
{\bf k}-representation we obtain $\epsilon({\bf k})=(1-2d)t({\bf k})$. 
Similarly, we find that $\epsilon_1({\bf k})=2dt({\bf k})$.
An analogous procedure is realized also in the equations for the other 
Green functions in~(4.2). 

Finally, in {\bf k}-representation the single-particle Green
function is
\begin{eqnarray}
&&G_{\bf k}(E)={1\over 2\pi}\left({A_{\bf k}\over E-
E_1({\bf k})}+{B_{\bf k}\over E-E_2({\bf k)}}\right),\\
&&A_{\bf k}={1\over 2}-{2dt({\bf k})\over \sqrt{U^2+(4dt({\bf k}))^2}}, \quad
B_{\bf k}={1\over 2}+{2dt({\bf k})\over \sqrt{U^2+(4dt({\bf k}))^2}}, \\
&&E_{1,2}({\bf k})=(1-2d)t({\bf k})\mp {1\over 2}
{\sqrt{U^2+(4dt({\bf k}))^2}}. 
\end{eqnarray}

 The single-particle Green function~(4.7) 
gives exact atomic and
band limits: if $U=0$ then $d=1/4$ and $G_{\bf k}(E)$ gets the band form, 
if $t({\bf k}) \rightarrow 0$ then we obtain exact atomic limit.

The energy gap (difference of energies between bottom of the upper and 
top of the lower Hubbard bands)
\begin{eqnarray}
\Delta E=-2w(1-2d)+\sqrt{U^2+(4dw)^2},
\end{eqnarray}
($w=z|t|,\ z$ is the number of nearest neighbours to a site). 

The distinction of Eq.~(4.7), (4.9)--(4.10) from earlier obtained 
results (see, for example [34, 35, 40]) is the dependence on the concentration of polar states 
(and thus on temperature). 

At $T=0$ the concentration of polar states is
\begin{eqnarray}
d=\left({1\over 4}+{U\over 32dw}\ln(1-4d)\right)\theta(2w-U),
\end{eqnarray}
($\theta(x)=1$ if $x>0$ and $\theta(x)=0$ if $x<0$).
$\Delta E\leq 0$ when the 
condition $2w\geq U$ is satisfied (in agreement with general physical 
ideas~[40]). In Fig.~2 the dependence of energy gap width on the ratio
$U/2w$ is shown.

 At $T\neq 0$ and given $w,\ U$ gap vanishes at $c<c_0$, where
\begin{eqnarray}
c_0={1-(U/2w)^2\over 4} \quad (2w>U);
\end{eqnarray}
if $c>c_0$ then insulating state is realized.
Thus, the proposed approach allows to describe melal-insulator transition.

\section{Specific narrow-band effects}
\setcounter{equation}{0}

\subsection{Absence of electron-hole asymmetry in NEB}

Let us consider narrow-band system in which the electron concentration 
$n<1$ and
the energy subbands $\sigma$-0 and $\uparrow\downarrow$-$\sigma$ are
separated by gap $\Delta{E}$. Thus at temperature $kT\ll\Delta{E}$ we can
limit ourselves by a consideration of the lower $\sigma-0$-subband. State 
of such system (doped Mott-Hubbard insulator -- DMHI) will be described
EH~(3.20) in which we take that the expressions coresponding to hopping 
$|\uparrow\downarrow>$-states are equal to a zero.

Let NEB is in the DMHI state with $n>1$. In the Hubbard model physical
properties of system of DMHI are equivalent both for $n<1$ and for $n>1$
when the condition $\langle{X}_i^0\rangle=\langle{X}_i^2\rangle$ is 
satisfied. This peculiarity of the Hubbard model (doublon-hole or electron-
hole symmetry) is a result of hopping integrals equality in $\sigma-0$- 
and $\uparrow\downarrow$-$\sigma$-subbands. In the proposed model hopping
integrals in both subbands $t_{ij}(n)$ and $\tilde{t}_{ij}(n)$ can be
esentially different, besides at the transition of system from the
state DMHI with $n<1$ to the state DMHI with $n>1$ bandwidth have the
jump equal to $2zT(ij)$ (and it continue to decrease with increase of $n$
in consequence of taking into account correlated hopping; see Fig.~3). 
So properties of
narrow-band system with strong intra-atomic interaction can be very 
different for cases $n<1$ and $n>1$ in consequence of the essential 
difference between subband widths (doublon-hole or electron-hole 
asymmetry).

This non-equivalence will be shown, in particular, in dependence of 
conductivity on degree of subband filling. In paper~[22] had been shown that
for DMHI conductivity at $n<1$ $\sigma\sim cnw/(2-n)$, 
and for $n>1$
$\tilde{\sigma}\sim d\tilde{w}(2-n)/n$, ($c=\langle{X}_i^0\rangle,
d=\langle{X}_i^2\rangle$). In the region of electron concentration for
which $\partial{\sigma}/\partial{n}>0(n<1)$ and
$\partial{\tilde\sigma}/\partial{n}>0(n>1)$ we have conductivity $n$-type,
for $\partial{\sigma}/\partial{n}<0)$, $\partial{\tilde\sigma}/\partial{n}<0$
-- conductivity $p$-type. One can see that $n-p$-type of conductivity of 
narrow-band system in the DMHI state is changed three time with change of
electron concentration from $0$ to $2$: in a region of first and second 
maximums (if we neglect correlated hopping then $n_1\simeq 0,6$ and 
$n_2\simeq 1,4$) and at $n=1$. In a region of some conductivity type the
expressions for calculation of conductivity can be written in the Drude-Lorentz 
form with effective mass depending on electron concentration~[22].

The non-equivalence of cases $n<1$ and $n>1$  in the concentration-dependence
of $\sigma(n)$ is confirmed experimentally. In the paper~[41] was shown 
that in metalooxides with less than half-filling of $3d$-shell (Œn$_2$Ž) 
conductivity is much higher than in the compounds with half or more than 
half-filling of $3d$-shell (ŒnŽ, NiŽ).

\subsection{An application of the model for a consideration of some
properties of narrow-band materials}

Let us shortly consider a possibility of application of the obtained 
results for the explanation of some narrow-band systems properties.

1. Binding energy of $3d$-metals. The binding energy in our model is
defined (for the case of weak and moderate intra-atomic interaction)
by the formula
\begin{eqnarray}
E_{b}=-\sum_{k\sigma} \epsilon_{{\bf k}\sigma}
\langle \alpha_{k{{\sigma}}}^{+}\alpha_{ k{{\sigma}}}\rangle-\nu{U},
\end{eqnarray}
where $\epsilon_{{\bf k}\sigma}$ -- Fourier-component of $t_{ij}$,
$\nu=n^2 /4$ for $n<1$ and $\nu=1-n+n^2 /4$ for $n>1$.
In the approximation of the rectangular density of states binding energy 
has the form
\begin{eqnarray}
E_{b}={1\over 2w(n)}\left[w^2(n)-t^2_c\right]-\nu U, 
\end{eqnarray}
with
\begin{eqnarray*}
w(n)=w_0\left[1-n(\tau_1+\tau_2) \right], \quad 
t_c=w(n)[n-1], 
\end{eqnarray*}
where $\tau_1,\ \tau_2$ are the parameters of correlated hopping,
$2w_0$ -- unperturbed bandwidth.
The dependence of binding energy on the $d$-electron concentration
in $3d$-systems can be determined by a generalization of Eq.~(5.1) for
the case of five equivalent $d$-subbands. Fig.~4 shows that the obtained 
results explain
the peculiarities of the dependence of binding energy on atomic number:
minimum for Œn and a being of two non-equivalent maximums (V, '®) 
(as the result of taking into account correlated hopping).

2. MIT under the action of external influences. From Eq.~(4.20) one can
see that energy gap increases with increase of current carrier concentration
at given $U$ and $w$ (i.e. at constant external pressure). This increase
can be caused by increase of temperature; the condition of metallic state
realization $c<c_0$ in this case can be not satisfied. The obtained
temperature dependence of $\Delta{E}$ can explain the observable 
transition in (V$_{1-x}$Cr$_x$)$_2$Ž$_3$ [42] ÷ NiS$_2$ [43] 
from paramagnetic metal to Mott-Hubbard insulator with increase
of temperature.

The dependence of $\Delta{E}$ on polar state concentration points out a
possibility of specific narrow-band effects giving an opportunity to
control MIT with the help of magnetic field and photoeffect.
For example, strong magnetic field can cause decrease of polar state
concentration $d$ [22] and MIT occurs. Otherwise, increase of  $d$ 
caused by photoeffect will stimulate the reverse metal-to-insulator
transition (analogous to temperature change).

3. Change of $n-p$ type of conductivity. Change of conductivity type 
about half-filling noted in Sect.~4 is confirmed experimentally for some
compounds, i.g. VŽ$_x$; in the frames of the considered model Mott-Hubbard
insulator state here at $å=0$ corresponds to the electron concentration
$n=1$ (modeling half-filled $t_2q$-band).  At $å>1$ in VŽ$_x$  holes
(V$^{3+}$) appear and at $x<1$ doublons (V$^{+}$) appear. 
In accordance with the result of Sect.~4 the experiment~[40] exhibit
at $å\simeq 1$ the transition from $p$-type conductivity (at $å>1$) 
to $n$-type conductivity (at $x<1$). Analogous change of conductivity
type is observed also in Co$_x$Fe$_{3-x}$O$_3$ [44].

4. Concentration-dependence of activation energy. In consequence of
concentration-dependence of the parameters in the quasiparticle energy
spectrum in $\sigma$-0 and $\uparrow\downarrow$-$\sigma$-subbands at the
transition of system from state with $n<1$ to state with $n>1$ activation
energy has a jump at $n=1$. In this case both increase and decrease
of activation energy are possible depending on mutual arrangement of 
$\sigma$-0 and $\uparrow\downarrow$-$\sigma$-subbands relative to other
bands. This jump of activation energy is confirmed experimentally for
Mn$_{x}$Fe$_{3-x}$Ž$_{4}$~[44] and '®$_{x}$Fe$_{3-x}$O$_{4}$~[41].

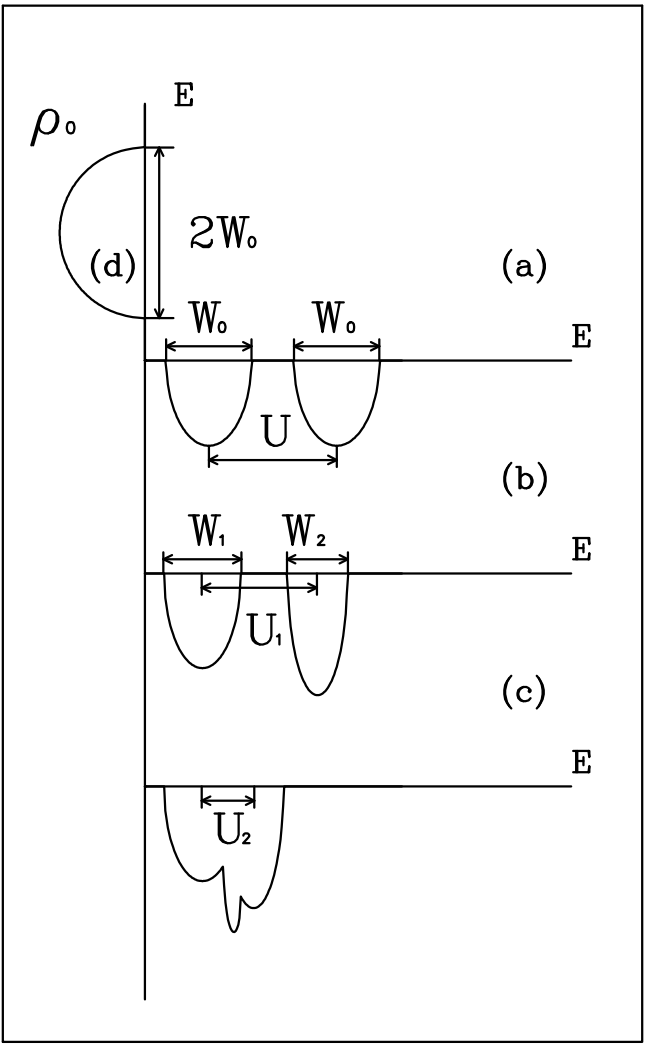
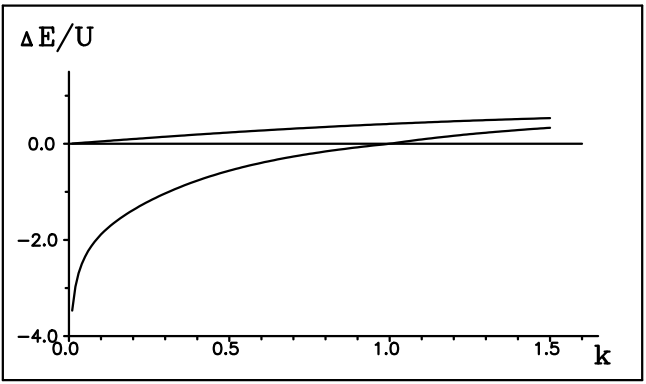
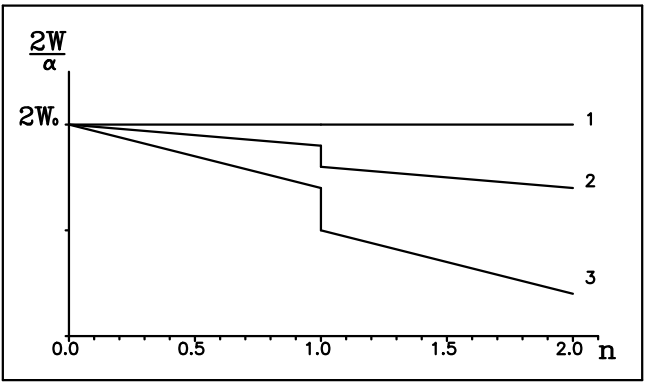
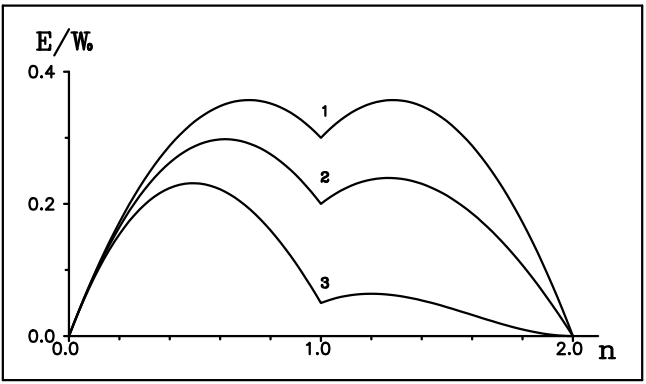
\end{document}